\documentclass[journal]{IEEEtran}

\usepackage{ifpdf}
\usepackage{graphicx}
\usepackage{amssymb}
\usepackage[tbtags]{amsmath}
\usepackage{amsmath}

\usepackage{cite}
\usepackage{epstopdf}
\usepackage{dsfont}
\usepackage[justification=centering]{caption}
\usepackage{color}
\usepackage{algorithm}
\usepackage{algorithmicx, algpseudocode}
\usepackage{etoolbox}
\usepackage{subcaption}

\usepackage{balance}

\newtheorem{theorem}{Theorem}

\makeatletter
\def\ScaleIfNeeded{%
\ifdim\Gin@nat@width>\linewidth \linewidth \else \Gin@nat@width \fi
} \makeatother

\newcommand{\mbf}[1]{\mathbf{#1}}
\begin{document}

\title{Dynamic Spectrum {\color{black} Sharing} for Load Balancing in Multi-Cell Mobile Edge Computing}



\author{Ming Zeng, \emph{Student Member}, \emph{IEEE}, and Viktoria Fodor, \emph{Member}, \emph{IEEE}
\thanks{M. Zeng and V. Fodor are with KTH Royal Institute of Technology, Stockholm, Sweden, e-mail: \{mzeng, vfodor\}@kth.se.}
\thanks{This work was supported in part by the Swedish Governmental Agency for Innovation Systems, under grant 2018-01554.}
}

\maketitle

\begin{abstract}
Large-scale mobile edge computing (MEC) systems require scalable solutions to allocate communication and computing resources to the users. In this letter we address this challenge by applying dynamic spectrum sharing among the base stations (BSs), together with local resource allocation in the cells.
We show that the network-wide resource allocation can be transformed into a convex optimization problem, and propose a distributed, hierarchical solution with limited information exchange among the BSs. Numerical results demonstrate that the proposed solution is superior to other baseline algorithms, when wireless and computing resource allocation is not jointly optimized, or the wireless resources allocated to the BSs are fixed.
\end{abstract}

\IEEEpeerreviewmaketitle

\section{Introduction}
By enabling mobile devices to offload computation-intensive tasks to servers in close proximity, mobile edge computing (MEC) can provide low-latency services for emerging applications, such as immersive augmented reality, wearable cognitive assistance, or autonomous driving.
{\color{black}Meanwhile, computation offloading} can decrease the energy consumption of the mobile devices \cite{BarbarossaSPM14}
and thus prolong their lifetime.

Early works on MEC focus on single cell systems with multiple users {\color{black} \cite{BarbarossaSPM14, Xchen, Ming_PIMRC}}.
Recently, the general scenario of multi-cell MEC is receiving attention \cite{S_TSP10, S_SIG13, C_Park18, Proa_INF19}. In \cite{S_TSP10}, a MIMO multicell system with a common edge server
is considered. The formulated energy minimization problem is solved using successive convex approximation. A game theoretic approach for the joint optimization of wireless and computing resources is proposed in \cite{S_SIG13}, while the performance of MEC in heterogeneous networks is  evaluated in \cite{C_Park18}, using stochastic geometry. A comprehensive study on the complexity of service placement and request routing in multi-cell MEC is provided in \cite{Proa_INF19}.
Most of the above works consider resource allocation in the multi-cell MEC as a large, centralized optimization problem, an approach that is not viable for large-scale systems.
Research on cellular networks faced the same issue, and provided the approaches of biasing (also called cell breathing) \cite{Ye_WCOM13, LinJSAC15}, and dynamic spectrum sharing (also called channel borrowing) \cite{KT_TN07, LMO_TVT08, Luo_JSAC08} to balance network traffic across the cells. Initial results for biasing in MEC are shown in \cite{C_Park18}.


In this letter we adapt dynamic spectrum sharing to achieve communication and computation load balancing among the BSs,
with the objective to minimize the total transmission energy consumption under computational delay constraints \cite{BarbarossaSPM14}.
We show that energy minimization can be transformed into a convex optimization problem, for which centralized optimal solution exists. Based on the centralized problem formulation, we propose a primal-dual resource allocation algorithm that {\color{black}lends} itself to an iterative distributed solution, where BSs cooperate to share the spectrum, while each individual BS allocates its local communication and computing resources to the associated users. Numerical results show that the joint resource allocation can reduce the energy consumption significantly, while the proposed distributed solution requires limited information exchange among the BSs and converges within a few iterations.

\section{System Model and Problem Formulation}
\label{systemmodel}
We consider a MEC system that consists of $K$ users, and $M$ BSs, each equipped with a MEC server. The users offload their computation tasks to a BS for processing.
We denote the set of users by $\mathcal{K} = \{1, \cdots, K\}$, and the set of BSs by $\mathcal{M} = \{1, \cdots, M\}$. We consider that each user $i \in \mathcal{K}$ generates computationally intensive {\color{black} and delay sensitive} tasks, characterized by {\color{black} three} parameters, the size $L_i$ of the input data, the number $W_i$ of CPU cycles required to perform the computation, and the completion time constraint $D_i$.

The objective of the considered MEC system is to minimize the energy consumption for data transmission under the delay constraint, by jointly allocating the wireless and computing resources, as well as the transmission power of the users.

{\bf{Communication resources:}}
The overall system bandwidth is $B$ Hz. We consider flat fading channel and orthogonal access with frequency division multiple access.  Users are associated to the BS with the best received signal-to-noise ratio, as it is often the case in today's cellular systems. Denote the corresponding channel gain for user $i$ by $h_i$. Then, the achievable data rate at user $i$ is given by
$R_i=x_i \log_2 \left( 1+\frac{P_i h_{i}}{x_i N_0} \right)$,
where $P_i$ is the corresponding transmission power, and $x_i$ denotes the allocated bandwidth, satisfying $\sum_{i \in \mathcal{K}} x_i =B$. Besides, $N_0$ is the noise power spectral density coefficient.
Accordingly, the transmission time and the resulting transmission energy consumption are respectively given by
$T_i=\frac{L_i}{R_i}$ and $E_i=\frac{L_i P_i}{R_i}$.

{\color{black}
Note that we consider orthogonal spectrum access here to reveal insights on joint resource allocation in MEC. Extension to multi-cell MEC systems with frequency reuse is discussed in Section V.
}

{\bf{Computing resources:}}
{\color{black} Let us denote the computational capacity of the MEC server at BS $j, j\in \mathcal{M}$ by $C_j$ and the set of users associated with BS $j$ by $\mathcal{S}_j, |\mathcal{S}_j|=K_j$.
The users served by the BS $j$, i.e., $\forall i  \in \mathcal{S}_j$ share the computing resource of the MEC server. We denote the computing resource allocated to user $i$ as $q_i$, satisfying $\sum_{i \in \mathcal{S}_j} q_i = C_j$.
Then, the computational time of user $i$'s task is given by $Q_i=\frac{W_i}{q_i}$ {\color{black}\cite{Liang_TWC19}}.


{\bf{Energy consumption minimization:}}
We consider the problem of total transmission energy minimization, under the constraint on the completion time of the computational tasks. That is, for each user $i$, the sum of the transmission and computational times should not
violate the maximum delay $D_i$, i.e., $T_i+Q_i \leq D_i$. The delay constraint then can be turned into the following rate requirement: $R_i \geq \frac{L_i}{D_i - Q_i}$.

The energy minimization problem can be formulated as
\begin{subequations} \label{P1}
\begin{align}
\text{P1}:& ~ \underset{\mbf{P},\mbf{x},\mbf{q}}{\text{min}} \sum_{i \in \mathcal{K}} E_i \\
\text{s.t.}~~ & ~R_i \geq \frac{L_{i}}{D_i-Q_i}, \forall i \in \mathcal{K} \\
& ~\sum_{i \in \mathcal{K}} x_{i} =B  \\
& ~ \sum_{i \in \mathcal{S}_j} q_i = C_j, \forall j \in \mathcal{M}
\end{align}
\end{subequations}
where $\mbf{P} \in \mathbb{R}^{K},\mbf{x} \in \mathbb{R}^{K},\mbf{q} \in \mathbb{R}^{K}$ are the vectors of allocated powers $P_{i}$, bandwidth $x_{i}$ and computational resource $q_i$, respectively.
Inequality constraints (\ref{P1}b) reflect the minimum data rate requirement for each user. Constraints (\ref{P1}c) limit the bandwidth, while (\ref{P1}d) restrict the computing resource.

\section{Centralized Resource Allocation}
To solve P1, the wireless and computing resources need to be allocated jointly. They are however coupled in a non-linear way through the delay constraint.
To progress with the solution, we first state the following theorem.

\begin{theorem}
Under any given bandwidth and computing resource allocation $\mbf{x},\mbf{q}$, the energy consumption is minimized when $T_i+Q_i=D_i$, $\forall i \in \mathcal{K}$ holds and the transmission power is set as
$ P_i=\frac{ N_0  x_i}{h_{i}} { \left(2^{\frac{R_i^{\min}}{x_i}}-1 \right)},
\forall i \in \mathcal{K}$
where $R_{i}^{\rm{min}}$ is the minimum rate that still fulfills the delay requirement, i.e., $R_{i}^{\rm{min}}=\frac{L_{i}}{D_i-Q_i}$.
\end{theorem}

\begin{IEEEproof}
When $x_i$ and $q_i$ are given, the energy consumption of the users is independent, and minimizing the total energy consumption is equivalent to minimizing that of each user. Without loss of generality, we look at $E_i$, which can be reformulated as
$ E_i=  \frac{L_i P_i }{R_i}=\frac{L_i P_i }{x_i \log_2 \left( 1+ \frac{P_i h_i}{x_i N_0} \right)}$.
Clearly, $E_i$ increases with $P_i$, and thereofre, $E_i$ is minimized when the minimum power is used. Meanwhile, to satisfy the delay constraint, we have $R_i= x_i \log_2 \left( 1+ \frac{P_i h_i}{x_i N_0} \right) \geq R_{i}^{\rm{min}}$, i.e., $P_i \geq {(2^{R_i^{\rm{min}}/x_i} -1)  N_0 x_i}/{ h_{i}}$. 
At equality the achieved rate is $R_{i}^{\rm{min}}$, which in turn results a transmission time of $T_i=D_i-Q_i$.
This concludes the proof.
\end{IEEEproof}

Let us then reformulate P1, based on Theorem 1. In addition, let us replace variables $q_i$ with
\begin{equation}
\label{eq:t-from-q}
t_i=D_i-{W_i}/{q_i}.
\end{equation}

%


This then leads to
\begin{subequations} \label{P2}
\begin{align}
\text{P2}:& ~ \underset{\mbf{x},{\mbf{t}}}{\text{min}} \sum_{i \in \mathcal{K}} \frac{  N_0 }{h_{i}} x_i t_i \left(2^{\frac{L_i}{x_i t_i}}-1 \right) \\
\text{s.t.}~~
& ~\sum_{i \in \mathcal{K}} x_{i} =B  \\
& ~ \sum_{i \in \mathcal{S}_j} \frac{W_i}{D_i-t_i} = C_j, \forall j \in \mathcal{M}
\end{align}
\end{subequations}

In P2, equality (\ref{P2}c) is clearly not affine, and thus, the feasible set is non-convex. 
{To address it, we relax the equality constraint and substitute (\ref{P2}c) with}
\begin{equation} \label{modified C2}
\sum_{i \in \mathcal{S}_j} \frac{W_i}{D_i-t_i} \leq C_j, \forall j \in \mathcal{M}
\end{equation}

As a consequence of Theorem 1, for any user $i$, the energy consumption decreases if $q_i$, the computing resource allocated to the user is increased. Thus, for the optimal solution, equality is achieved in \eqref{modified C2}, which means substituting (\ref{P2}c) with \eqref{modified C2} will not change the solution.

\begin{theorem}
Problem P2 with the relaxed constraint \eqref{modified C2} is a convex optimization problem.
\end{theorem}

\begin{IEEEproof}
First, equality constraint (\ref{P2}b) is affine. Then, for inequality constraint \eqref{modified C2}, its second derivative is $\sum_{i \in \mathcal{S}_j} \frac{2 W_i}{(D_i -t_i)}>0$, and thus, it is convex. Last, let us consider the objective function (\ref{P2}a). It can be seen that the energy consumption for each user is only affected by its own variables, e.g., for user $i$, $N_0  x_i t_i \left(2^{\frac{L_i}{x_i t_i}}-1 \right)/{h_{i}}$ is only affected by $x_i$ and $t_i$. Therefore, we can consider each user separately. Without loss of generality, we consider user $i$, whose Hessian matrix is given by {\color{black}
\[
\mbf{H}_i =
  \frac{N_0}{h_i} \cdot \begin{bmatrix}
    \mbf{H}_i (1,1)
     & \mbf{H}_i (1,2) \\
    \mbf{H}_i (2,1)
     & \mbf{H}_i (2,2)
  \end{bmatrix},
\]
where $\mbf{H}_i (1,1)=\ln2^2 \cdot 2^{\frac{L_i}{ t_i x_i} } \cdot \frac{L_i^2}{t_i x_i^3}$, while $\mbf{H}_i (2,2)=\ln2^2 \cdot 2^{\frac{L_i}{x_i t_i} } \cdot \frac{L_i^2}{x_i t_i^3}$. Besides, $\mbf{H}_i (1,2)=\mbf{H}_i (2,1)=2^{\frac{L_i}{x_i t_i} }-1 + \ln2^2 \cdot \frac{L_i^2}{x_i^2t_i^2} 2^{\frac{L_i}{x_i t_i} }- \ln 2 \cdot \frac{L_i}{x_i t_i} 2^{\frac{L_i}{x_i t_i} }$.
After some algebraic manipulations, it can be verified that $\rm{det}(\mbf{H}_i)>0$ holds for all $\frac{L_i}{x_i t_i}>0$, which indicates (\ref{P2}a) is convex. This completes the proof.}
%
\end{IEEEproof}

Based on Theorem 2, the optimal solution of P2 can be obtained using standard convex optimization methods in a centralized manner.

\section{Distributed Resource Allocation with dynamic Spectrum Sharing}
In this section we propose an Iterative Resource Allocation algorithm to solve problem P2, that lends itself to a distributed implementation, with decreased signaling needs.
As shown in \text{Algorithm \ref{alg:iterative}}, it follows two iterative steps: i)  {\color{black}the Bandwidth Allocation Algorithm (BAA)
updates $\mbf{x}$ to allocate bandwidth across and within the BSs, for given $\mbf{t}$,  and ii) the Computation resource Allocation Algorithm (CAA) updates $\mbf{t}$ to allocate the computing resource at each BS, for given bandwidth allocation $\mbf{x}$.}
We denote by $E_i^{t}$ and $E_i^{x}$ the energy consumption of user $i$ after optimizing $t_i$ and $x_i$, respectively, and $\epsilon$ is the stop condition.

\begin{algorithm}
\caption{Iterative Resource Allocation}
\label{alg:iterative}
\begin{algorithmic}[1]
\State {\textbf{Initialization:}}
 $q_i \leftarrow C_j/K_j, t_i \leftarrow \left(D_i-\frac{W_i}{q_i} \right),
\forall i \in \mathcal{S}_j, j \in \mathcal{M}$;
\State Update  $x_i, \forall i \in \mathcal{K}$ {\color{black}based on} BAA, and calculate $\sum\limits_{i} E_i^x$;
\State $\sum_{i} E_i^t \leftarrow \sum_{i} E_i^x+ 2 \epsilon$;
\State \textbf{while} $\sum_{i} E_i^{t} -  \sum_{i} E_i^{x} > \epsilon$ \textbf{do}
\State \hspace{20pt} Update $t_i$ based on CAA, and recalculate $\sum_{i} E_i^t$;
\State \hspace{20pt} Update $x_i$ based on BAA, and recalculate $\sum_{i} E_i^x$;
\State \textbf{end while}
\end{algorithmic}
\end{algorithm}


{\color{black}
{\bf{The Bandwidth Allocation Algorithm (BAA):}}
Assuming that the computing resource allocation $\mbf{t}$ is given, problem P2 is simplified as
\begin{subequations} \label{P3}
\begin{align}
\text{P3}:  ~ \underset{\mbf{x}}{\text{min}}~  \sum_{i \in \mathcal{K}} \frac{  N_0 }{h_{i}} t_i  x_i  \left(2^{\frac{L_i}{t_i x_i}}-1 \right)
~\text{s.t.}~ (\ref{P2}\rm{b}).
\end{align}
\end{subequations}

Since $\mbf{H}_i(1,1)>0$, P3 is a convex problem, and we can use the Karush-Kuhn-Tucker (KKT) condition to derive the optimal $\mbf{x}$.
The KKT condition for user $i$ is
\begin{align*} \label{lambda}
g(x_i)= \frac{ N_0 t_i}{h_i} \left[ 2^{\frac{L_i}{t_i x_i}} -    \frac{ L_i}{t_i x_i}  2^{\frac{L_i}{t_i x_i} } \ln2-1 \right] + \lambda =0, \forall i \in \mathcal{K}
\end{align*}
where $\lambda$ is the introduced auxiliary variable, satisfying $\lambda>0$. For given $\lambda$, the above equation can be used to obtain $x_i$. Specifically, we have $\frac{\partial g(x_i)}{\partial x_i} = \frac{ \ln 2^2 \cdot N_0 L_i^2}{h_i t_i x_i^3}   2^{\frac{L_i}{t_i x_i}}>0$, which indicates that $g(x_i)$ grows with $x_i$, and thus a bisection search can be used to obtain $x_i$ by comparing $g(x_i)$ with 0.
Now the problem lies in how to obtain $\lambda$. When $\lambda$ is increased, $x_i, \forall i \in \mathcal{K}$ will decrease to ensure $g(x_i)=0$. Meanwhile, $\sum_i x_i= B$ needs to hold. Consequently, $\lambda$ can also be obtained with bisection search, by comparing $\sum_i x_i$ with $B$.}

The resulting \color{black}BAA} consists of two loops: an outer loop to find the value of $\lambda$ and an inner loop to determine the bandwidth allocation ${\bf{x}}$.



\begin{figure*} \label{fig_1}
\centering
\begin{subfigure}{0.33\textwidth}
  \centering
  \includegraphics[width=1\linewidth]{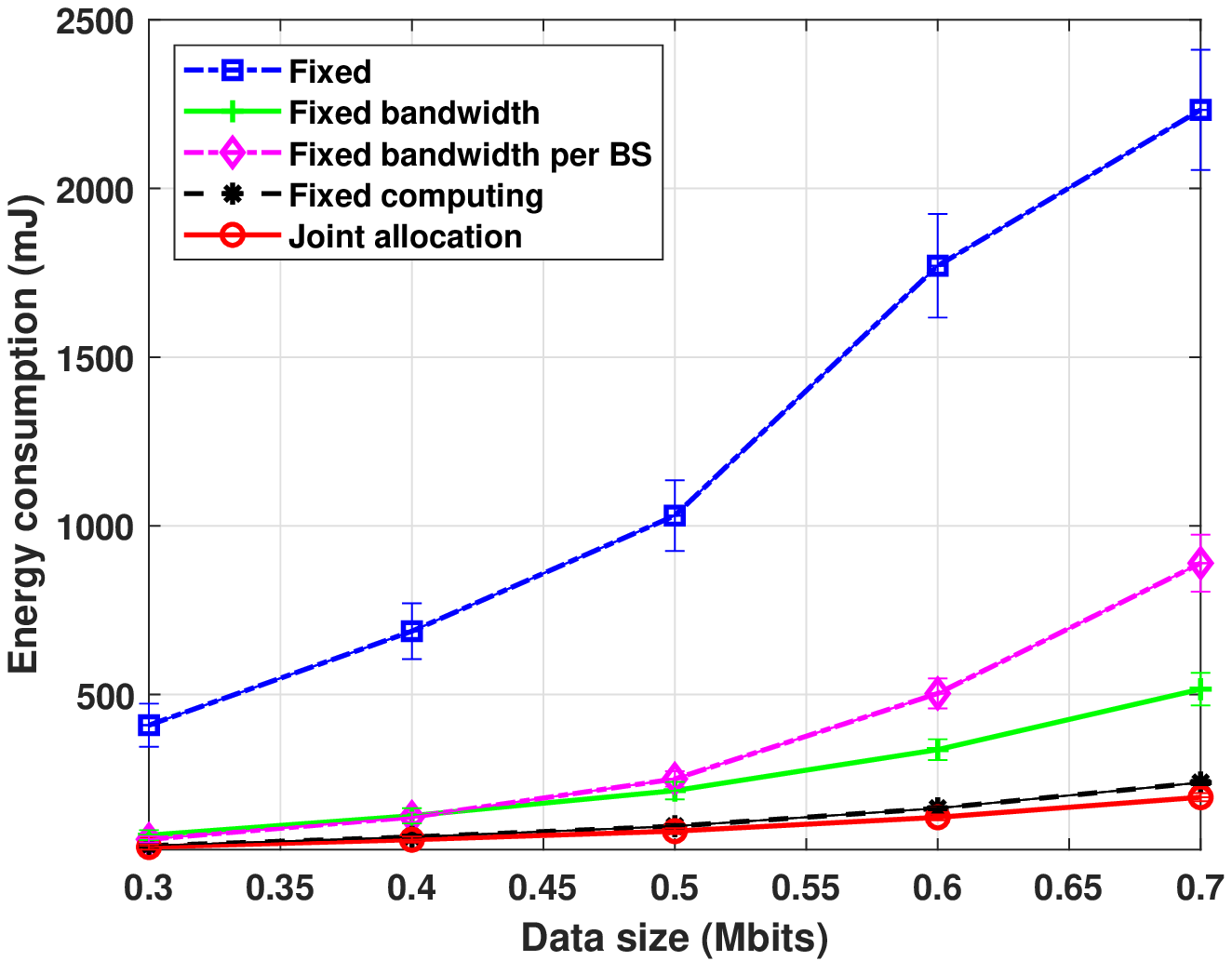}
  \caption{}
\end{subfigure}%
\begin{subfigure}{0.33\textwidth}
  \centering
  \includegraphics[width=1\linewidth]{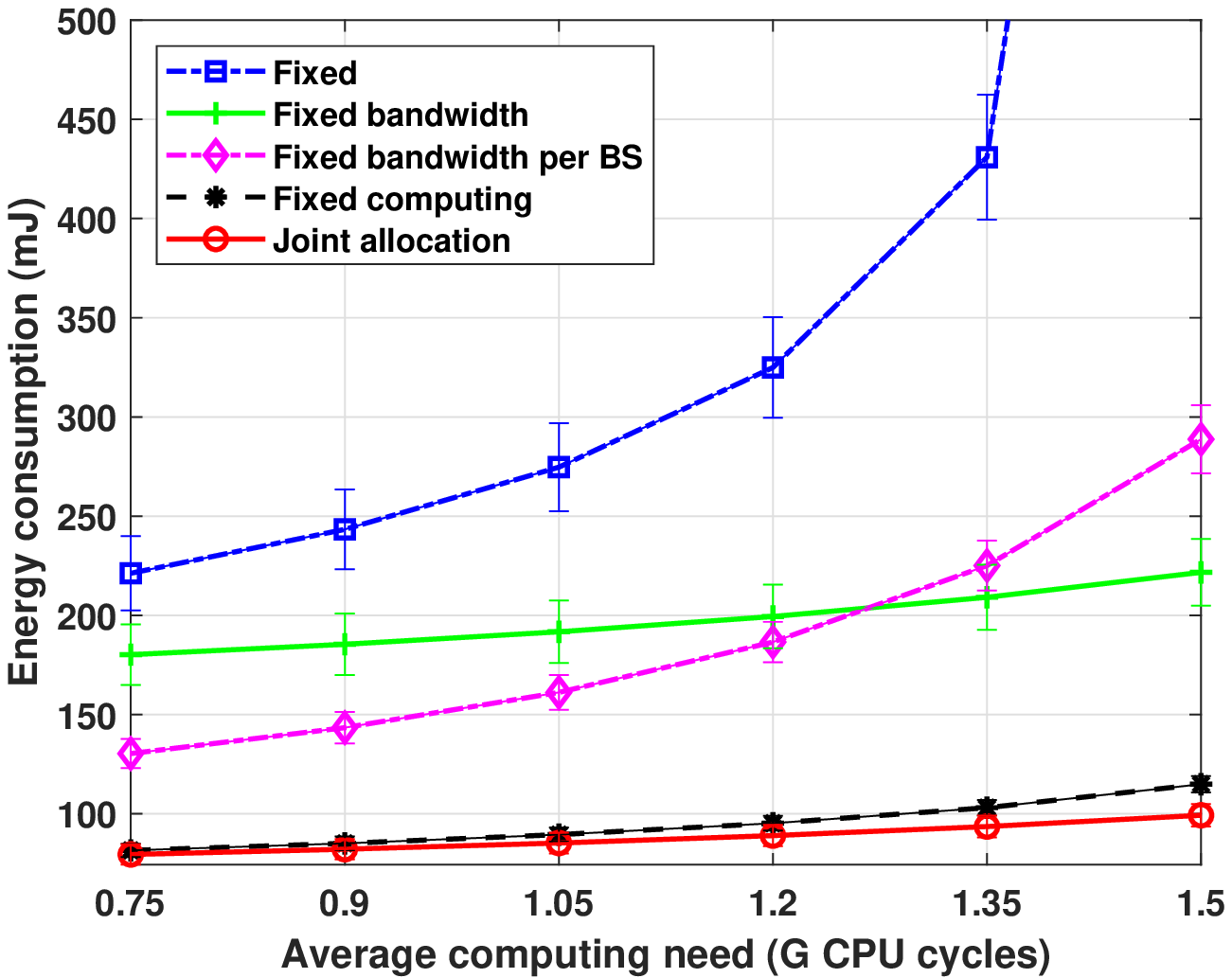}
  \caption{}
\end{subfigure}
\begin{subfigure}{0.33\textwidth}
  \centering
  \includegraphics[width=1\linewidth]{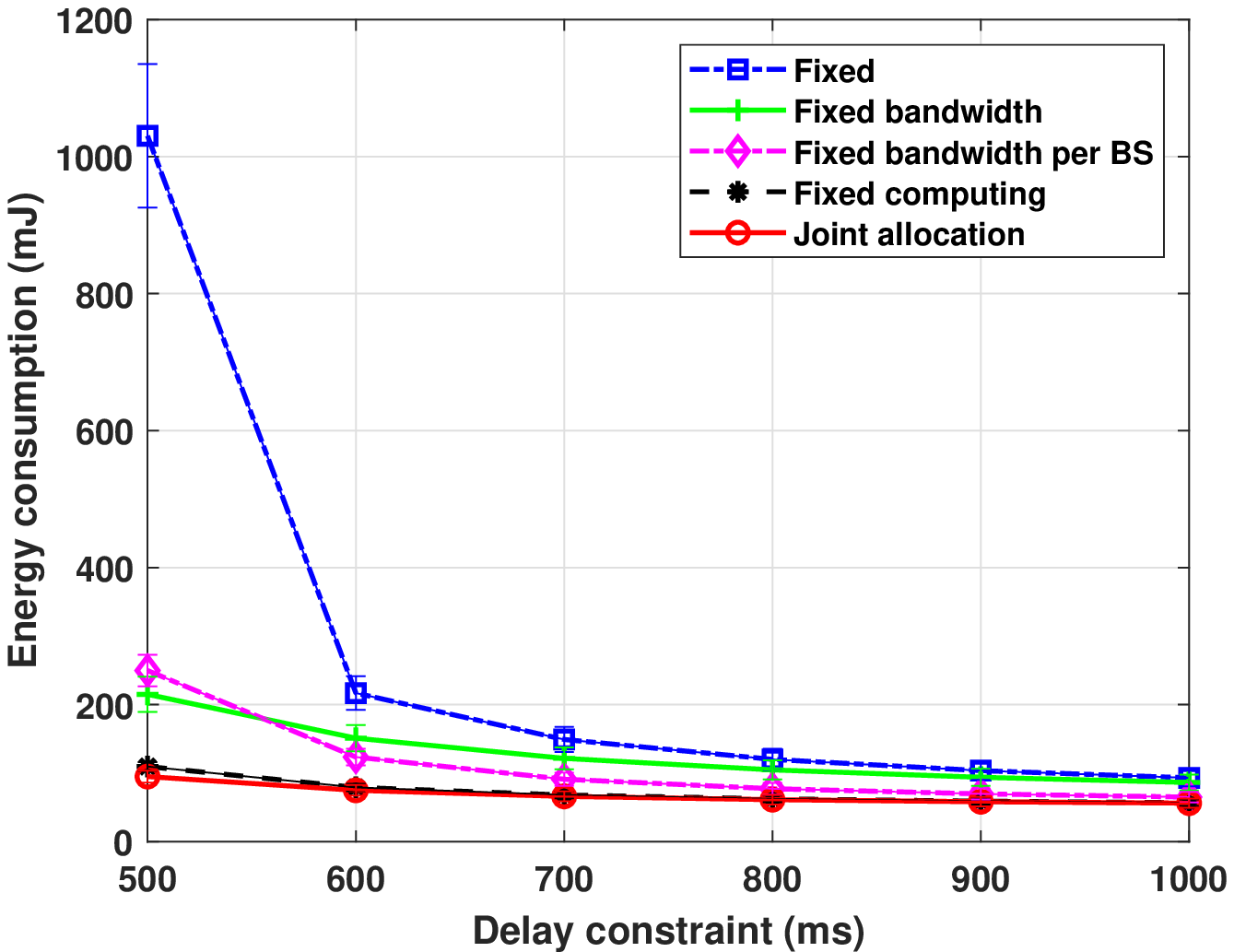}
  \caption{}
\end{subfigure}
\caption{Energy consumption as a function of (a) the data size, {\color{black}(b) the average computing need}, and (c) the delay constraint.}
\end{figure*}

{\bf{The Computing resource Allocation Algorithm (CAA):}}
Under given bandwidth allocation, the computing resource allocation is independent across the BSs. Thus, the energy minimization for each BS is equivalent to that of the overall system. 
{\color{black} Let us} consider BS $j$ and user set $\mathcal{S}_j$, {\color{black}$j\in \mathcal{M}$}. The corresponding optimization problem can be formulated as
\begin{align}
\text{P4}: ~ \underset{\mbf{t}}{\text{min}} \sum_{i \in \mathcal{S}_j} \frac{  N_0  }{h_{i}}  x_i t_i \left(2^{\frac{L_i}{x_i t_i}}-1 \right)
~\text{s.t.}~ \eqref{modified C2}.
\end{align}

As P3, P4 is also a convex problem, and the KKT condition is given by
\begin{align*}
&f(t_i)= \frac{ N_0 x_i}{ h_i} \left[ 2^{\frac{L_i}{x_i t_i}} -   \frac{L_i}{x_i t_i} 2^{\frac{L_i}{x_i t_i}} \ln2 -1 \right] + \frac{W_i }{(D_i-t_i)^2} {\color{black}\mu_j}=0, \\
&{\color{black}\forall i\in \mathcal{S}_j, j\in \mathcal{M}}
\end{align*}
where ${\color{black}\mu_j}$ is the introduced auxiliary variable, satisfying ${\color{black}\mu_j} \geq 0$.

Since $\frac{\partial f(t_i)}{\partial t_i}=\frac{ \ln 2^2 \cdot N_0 L_i^2}{h_i x_i t_i^3}   2^{\frac{L_i}{x_i t_i}} + \frac{4W_i }{(D_i-t_i)^2} {\color{black}\mu_j} >0$, we can conclude that $f(t_i)$ grows with $t_i$, and further $t_i$ declines with ${\color{black}\mu_j}$. Therefore, $t_i$ and ${\color{black}\mu_j}$ can be found with bisection search.
{\color{black}At each BS $j, j\in \mathcal{M}$,} the resulting {\color{black}CAA} includes an outer loop to find the value of ${\color{black}\mu_j}$ and an inner loop to determine ${\bf{t}}$, which in turn gives the computing resource allocation ${\bf{q}}$, according to \eqref{eq:t-from-q}.



%

{\bf{Distributed Implementation with Dynamic Spectrum Sharing among the BSs:}}

The Iterative Resource Allocation algorithm requires the implementation of BAA and CAA, and the exchange of the parameters between these algorithms. Note that CAA can be performed by the individual BSs. Similarly, for BAA, the update of $x_i$ under given $\lambda$ can happen locally at the BS. Finding the appropriate $\lambda$ value for the KTT condition however requires collaboration. Specifically, the BSs need to share their $\sum_{i \in \mathcal{S}_j}{x_i}$ values, that is, the bandwidth that should be allocated to BS $j$, and increase or decrease $\lambda$ in the bisection search, if $\sum_{j \in \mathcal{M}}\sum_{i \in \mathcal{S}_j}{x_i}$ is larger or smaller than $B$.

{\bf{Optimality and complexity:}}
\begin{theorem}
The Iterative Resource Allocation algorithm gives the optimal resource allocation in finite steps, with predefined accuracy $\epsilon$.
\end{theorem}

\begin{IEEEproof}
In both lines 5 and 6 of Algorithm \ref{alg:iterative}, the energy consumption decreases, or remains unchanged. Since there is a lower bound for the energy consumption, e.g., 0, the Iterative Resource Allocation algorithm always terminates, either by reaching the lower bound, or by achieving a decrease less than $\epsilon$. Moreover, the obtained local optimum is also the global optimum since the considered problem is convex.
\end{IEEEproof}

%
%


The centralized implementation requires the collection of user parameters and the distribution of the resource allocation vectors to the BSs, thus, the signaling complexity is $\mathcal{O}(K)$, where $K$ is the total number of users in the multi-cell system. The computational complexity comes form the iterations of Algorithm \ref{alg:iterative}, where both BAA and CAA perform bisection search for $\lambda$ and $\mu$ as well as for the $x_i$ and $t_i$ values. This gives a computational complexity of $\mathcal{O}(NK)$, where $N$ is the number of iterations in Algorithm \ref{alg:iterative}.

The distributed implementation requires information exchange among the BSs, to search for $\lambda$ in BAA, in each iteration steps of Algorithm \ref{alg:iterative}. {\color{black}This leads to a signaling overhead} of $\mathcal{O}(N M)$, where $M$ is the number of BSs.
{\color{black}Each BS $j$ needs to run BAA and CAA locally, and thus, the computation complexity is $\mathcal{O} (NK_j)$.}

The distributed implementation has good scalability properties, however, the complexity depends on the number of iterations $N$. Therefore, in \text{Section \ref{sec:num}} we investigate how $N$ depends on the network parameters.

{\color{black}\section{{\color{black}Multi-cell MEC with Frequency Reuse}}}
{\color{black} The previous sections consider orthogonal spectrum allocation among cells to reduce the complexity of the analysis and reveal insights.}
{\color{black}Frequency reuse is {\color{black} however necessary in large systems} to increase spectrum efficiency. {\color{black} To this end, let us first consider the case when fixed frequency reuse (i.e., according to 3 or 7 cell pattern) is adopted to avoid co-channel interference. In this case,
we can reformulate P2 as}{\color{black}
\begin{subequations} \label{P2_new}
\begin{align}
\text{P5}:& ~ \underset{\mbf{x},{\mbf{t}}, B_f}{\text{min}} \sum_{i \in \mathcal{K}} \frac{  N_0 }{h_{i}} x_i t_i \left(2^{\frac{L_i}{x_i t_i}}-1 \right) \\
\text{s.t.}~~
& ~\sum_{i \in \mathcal{S}_j} x_{i} =B_f, \forall j \in \mathcal{M}_f, f = \{1, \cdots, F\}  \\
& ~\sum_{f =1}^F B_f =B \\
& ~ \sum_{i \in \mathcal{S}_j} \frac{W_i}{D_i-t_i} = C_j, \forall j \in \mathcal{M}
\end{align}
\end{subequations}
where $F$ denotes the cell reuse factor, and $\mathcal{M}_f$ represents the cell set using the same frequency band $B_f$, $f = \{1, \cdots, F\}$.
Then, (\ref{P2_new}b) denotes the bandwidth constraint for each cell, while (\ref{P2_new}c) is the total bandwidth constraint. Both (\ref{P2_new}b) and (\ref{P2_new}c) are affine constraints, and thus problem P5 with a relaxed (\ref{P2_new}d) (i.e., \eqref{modified C2}) is convex, and can be easily solved using standard convex optimization tools. }

{\color{black} An iterative solution that also allows distributed implementation can follow the lines of Algorithm 1. CAA can be performed as described in Section IV, but the bandwidth allocation algorithm has to be extended.}
{\color{black} 
Now the KKT conditions are given by
\begin{align} \label{lambda_j}
&\frac{ N_0 t_i}{h_i} \left[ 2^{\frac{L_i}{t_i x_i}} -    \frac{ L_i}{t_i x_i}  2^{\frac{L_i}{t_i x_i} } \ln2-1 \right] + \lambda_j =0, \forall i \in \mathcal{S}_j,  j \in \mathcal{M}  \\
&\beta =\sum_{j \in \mathcal{M}_f}  \lambda_j, \forall f = \{1, \cdots, F\} \label{lambda_j_2}
\end{align}
where $\lambda_j$ and $\beta$ are the introduced auxiliary variables for (\ref{P2_new}b) and (\ref{P2_new}c), respectively, satisfying $\lambda_j, \beta>0$.
}

{\color{black} Algorithm 2 summarizes the steps to find {\color{black}$B_f$}, $x_i$, $\lambda_j$ and $\beta$. The algorithm has an inner loop to determine {\color{black}$B_f$}, $x_i$ and $\lambda_j$ for given $\beta$, according to BAA in Section IV and {\color{black}\eqref{lambda_j_2}}. This iteration ensures that the bandwidth is optimally allocated for given {\color{black}$\beta$} values. Then, an outer loop finds $\beta$, such that constraint (\ref{P2_new}c) is satisfied.
The distributed implementation requires $\beta, B_f$ and $\lambda_j $ to be exchanged among the cells.}

Now let us consider the extreme case with universal frequency reuse. Due to the existence of co-channel interference, users' achievable rates are non-convex functions over their transmit powers.
As a result, the energy minimization problem is likely to be NP-hard \cite[Theorem 1]{Luo_JSAC08}.
To make it tractable, we may need to refer to convex approximation or dual optimization \cite{Luo_JSAC08, S_TSP10, Ywei_TCOM06}.

\begin{algorithm}
\caption{Bandwidth Allocation with Frequency Reuse}\label{BM}
\label{alg:bisec}
\begin{algorithmic}[1]
\State {\textbf{Initialization:}} $\beta_{\rm{low}}$; $\beta_{\rm{up}}$; $\epsilon$
\State \textbf{while} $ \beta_{\rm{up}}- \beta_{\rm{low}} > \epsilon$
\State \hspace{15pt} $\beta \leftarrow \frac{\beta_{\rm{low}} + \beta_{\rm{up}}}{2}$;
\State \hspace{15pt}  \textbf{for} $f \leftarrow 1, \cdots, F$
\State \hspace{30pt}  initialization: $B_f^{\rm{low}} $, $B_f^{\rm{up}} $;
\State \hspace{30pt} \textbf{while} $ B_f^{\rm{up}}- B_f^{\rm{low}} > \epsilon$
\State \hspace{45pt}  $B_f \leftarrow \frac{B_f^{\rm{low}} + B_f^{\rm{up}}}{2}$;
\State \hspace{45pt}  obtain $x_i, \lambda_j, i \in \mathcal{S}_j, j\in \mathcal{M}_f$ as in BAA;
\State \hspace{45pt}  \textbf{if} $\sum_{j \in \mathcal{M}_f} \lambda_j> \beta$ then $B_f^{\rm{low}} \leftarrow B_f$;
\State \hspace{45pt}  \textbf{else} $B_f^{\rm{up}} \leftarrow  B_f$;
\State \hspace{45pt} \textbf{end};
\State \hspace{30pt} \textbf{end while};
\State \hspace{15pt} \textbf{end for};
\State \hspace{15pt}  \textbf{if}  $\sum_{f =1}^F B_f>B$ then $\beta_{\rm{low}} \leftarrow  \beta$
\State \hspace{15pt} \textbf{else} $\beta_{\rm{up}} \leftarrow  \beta$
\State \hspace{15pt} \textbf{end};
\State  \textbf{end while}
\end{algorithmic}
\end{algorithm}
}

\begin{figure*} \label{fig_2}
\centering
\begin{subfigure}{0.33\textwidth}
  \centering
  \includegraphics[width=1\linewidth]{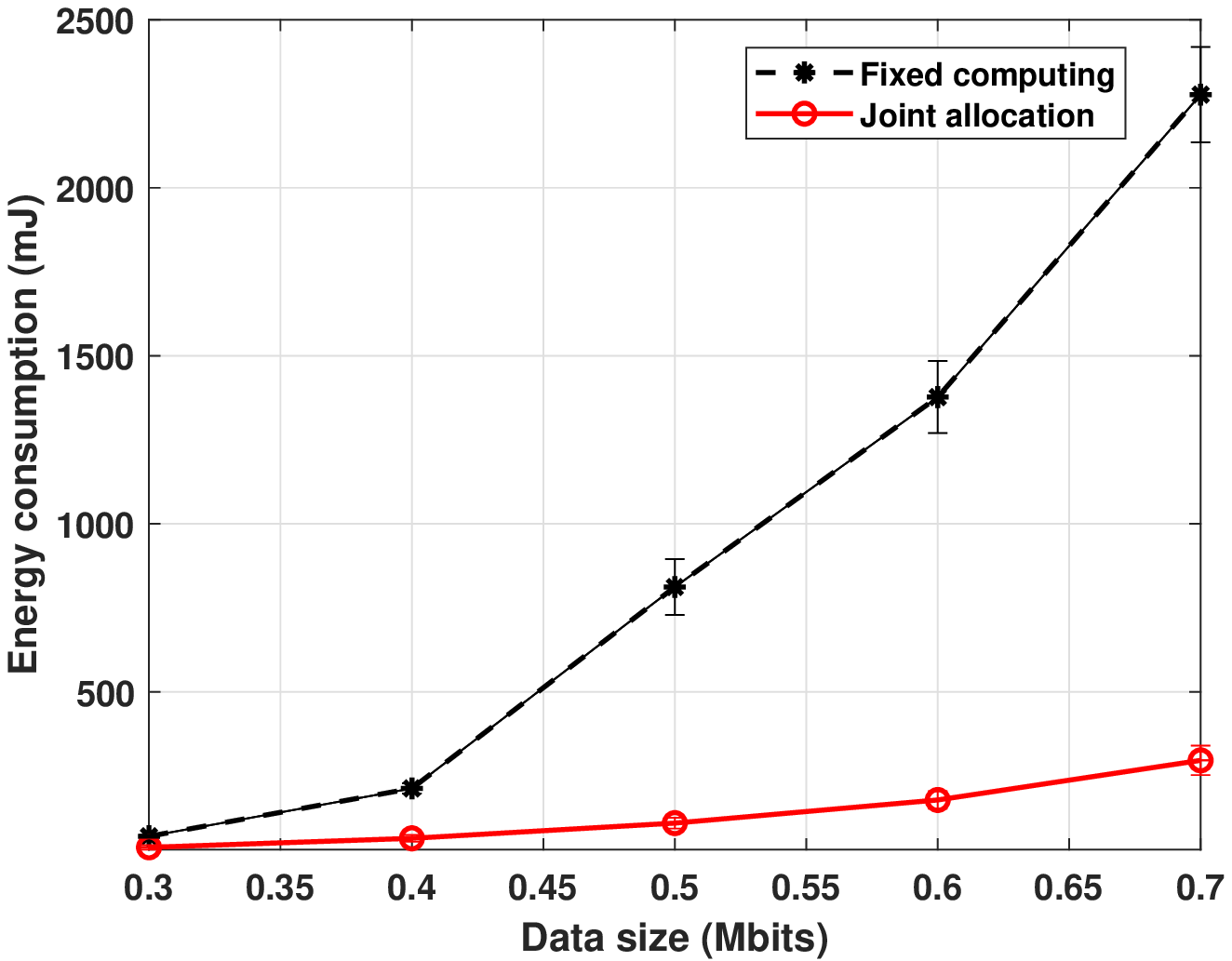}
  \caption{}
\end{subfigure}%
\begin{subfigure}{0.33\textwidth}
  \centering
  \includegraphics[width=1\linewidth]{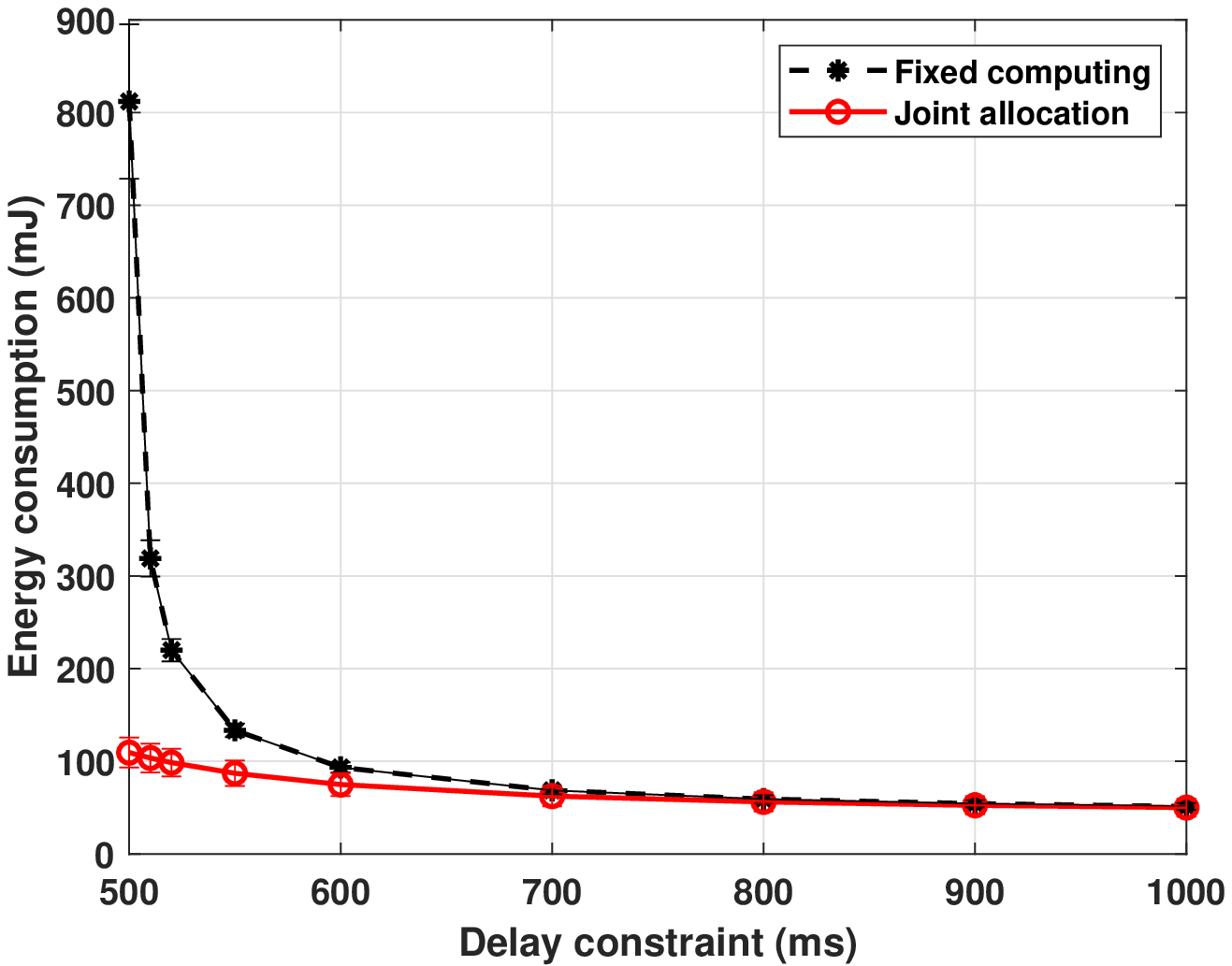}
  \caption{}
\end{subfigure}
\caption{{\color{black}Energy consumption for Fixed computing and Joint allocation as a function of (a) data size, and (b) delay constraint.}}
\end{figure*}
\section{Numerical Results} \label{sec:num}

We evaluate the performance of the joint bandwidth and computing resource allocation scheme in a simulator implemented in Matlab. For each trial, we place the BSs and the users randomly uniformly in a disk of a radius of $200$ m. The pathloss model follows $30.6+36.7\log_{10}(d)$, where $d$ is the distance in m. Rayleigh fading is used for small-scale fading. We set $B=10$ MHz, $N_0=-174$ dBm/Hz and $\epsilon = 10^{-6}$.

We consider four baseline algorithms: a) equal bandwidth and computing resource per user, referred to as Fixed;  b) equal bandwidth per user, while the computing resource is optimized, referred to as Fixed bandwidth; c) equal bandwidth for each BS, but optimized joint resource allocation within each BS, referred to as Fixed bandwidth per BS;  and d) equal computing resource per user, with optimized bandwidth, referred to as Fixed computing.

Fig. 1 shows the energy consumption under the five algorithms when the data size, the {\color{black} average} computing need, and delay constraint vary, respectively. The default simulation values are: $M=4$, $K=32$, $C_j=100$ G CPU cycles/s, $L_i=0.5$ Mbits, $D_i=500$ ms. For Figs 1(a) and (c), $W_i$ is generated randomly uniformly within $[0.5,2.5]$ G CPU cycles for each user. 
{\color{black}In Fig. 1(b), $W$ is increased, and for each user $W_i$ is generated randomly uniformly within $[\frac{1}{3}W, \frac{5}{3}W]$.}

As expected, the energy consumption grows with the data size and computing need, but decreases as delay constraint gets relaxed. The proposed Joint resource allocation  always achieves the best performance. The large difference between Joint allocation and Fixed bandwidth illustrates the gain of optimizing the bandwidth allocation among the users. Likewise, the difference between Joint allocation and Fixed bandwidth per BS indicates that the load in the cells can be highly unbalanced, and thus dynamic bandwidth sharing among the BSs is necessary.
{\color{black}Fixed computing has similar performance to Joint allocation, the reason is probably that 
the disparity among users' computing needs is small in the considered scenario.
Therefore, in Fig. 2 we present the corresponding results
with a higher variance, i.e., $W_i$ is generated randomly uniformly within $[0.5, 4]$ G CPU cycles for each user.
It can be seen that Joint allocation consumes much lower energy than Fixed computing, especially under large data size or strict delay constraint.}

We also conducted extensive simulations to evaluate $N$, the number of iterations required for the Iterative Resource Allocation algorithm to converge. We found that $N$ does not depend significantly on $M$, the number of BSs, for example, for $M=16$ and $K=64$ the algorithm converges in two iterations on average. However, $N$ increases almost linearly with $K_j$, the number of users in a cell. {\color{black}For example, under $M=4$, the average number of iterations increases from two to four when $K_j$ changes from $K_j = 8$ to $K_j = 16$.}

\section{Conclusion}
In this paper, we considered a multi-user multi-cell MEC system, where users offload their computing tasks to the BS {\color{black}with the best channel} for processing.
An overall transmission energy minimization problem was formulated
and transformed into a convex optimization problem. Furthermore, a scalable distributed solution was proposed inspired by the dynamic spectrum sharing approach in cellular networks. Numerical results showed that the proposed joint allocation outperforms other baseline algorithms, when wireless and computing resources are not jointly optimized, or the wireless resources allocated to the BSs are fixed.


\balance

\bibliographystyle{IEEEtran}
\bibliography{IEEEabrv,conf_short,jour_short,mybibfile}

\end{document}